\documentclass[12pt]{article}

\usepackage{amssymb}
\usepackage{amsthm}
\usepackage{amsmath}
\usepackage{booktabs}
\usepackage{graphicx}
\usepackage{natbib}
\usepackage[hyphens]{url}
\usepackage{color,soul}
\usepackage{multirow}
\usepackage[top=1in, bottom=1in, left=1in, right=1in]{geometry}
\usepackage[linesnumbered,ruled]{algorithm2e}
\usepackage[labelformat=simple]{subcaption}

\newcommand{\clr}{\operatorname{clr}}
\newcommand{\alr}{\operatorname{alr}}
\newcommand{\ilr}{\operatorname{ilr}}

\newcommand{\var}{\operatorname{Var}}
\newcommand{\bfx}[1]{{\bf #1}}

\title{Compositional Data Regression in Insurance with Exponential Family PCA}

\author{Guojun Gan\thanks{Corresponding author, Department of Mathematics, University of Connecticut, 341 Mansfield Road, Storrs, CT, 06269-1009, USA. Email: \texttt{guojun.gan@uconn.edu}.} \and Emiliano A. Valdez\thanks{Department of Mathematics, University of Connecticut, 341 Mansfield Road, Storrs, CT, 06269-1009, USA. Email: \texttt{emiliano.valdez@uconn.edu}.}}

\begin{document}

\maketitle

\begin{abstract}
Compositional data are multivariate observations that carry only relative information between components. Applying standard multivariate statistical methodology directly to analyze compositional data can lead to paradoxes and misinterpretations. Compositional data also frequently appear in insurance, especially with telematics information. However, such type of data does not receive deserved special treatment in most existing actuarial literature. In this paper, we explore and investigate the use of exponential family principal component analysis (EPCA) to analyze compositional data in insurance. The method is applied to analyze a dataset obtained from the U.S. Mine Safety and Health Administration. The numerical results show that EPCA is able to produce principal components that are significant predictors and improve the prediction accuracy of the regression model. The EPCA method can be a promising useful tool for actuaries to analyze compositional data.


\end{abstract}

\newpage

\section{Introduction}

Compositional data refer to data that describe parts of some whole and are commonly presented as vectors of proportions, percentages, concentrations, or frequencies. The sums of these vectors are constrained to be some fixed constant, e.g. 100\%.  Due to such constraints, compositional data possess the following properties~\citep{pawlowsky2015modelling}: 
\begin{description}
	\item \textbf{Scale invariance}. They carry relative rather than absolute information. In other words, the information does not depend on the particular units used to express the compositional data.
	\item \textbf{Permutation invariance}.  Permutation of components of a compositional vector does not change the information contained in the compositional vector.
	\item \textbf{Subcomposition coherence}. Information conveyed by a compositional vector of $d$ components
	should not contradict that coming from a subcomposition containing $d_\ast (d_\ast <d)$ components of the original $d$ components.
\end{description}
These properties are important to understand the consequences that can make the application of conventional statistical methods restraint and invalid for compositional data.

Consider the case that such data are multivariate observations that carry relative information between components.  Applying standard statistical methodology designed for unconstrained data directly to analyze such compositional features can affect inference leading to paradoxes and misinterpretations. Table \ref{tbl:simpson} shows an example of Simpson's paradox \citep{pawlowsky2015modelling}. Table \ref{tbl:simpson1} shows the number of students in two classrooms who arrive on time and are late, classified by gender. Table \ref{tbl:simpson2} shows the corresponding proportions of students arriving on time or being late. Table \ref{tbl:simpson3} shows the aggregated number of students from the two classrooms who arrive on time and are late. The corresponding proportions are shown in Table \ref{tbl:simpson4}. From Table \ref{tbl:simpson2}, we see that the proportion of female students arriving on time is greater than that of male students in both classrooms. However, Table \ref{tbl:simpson4} shows the opposite result: the proportion of male students arriving on time is greater than that of female students. The Simpson's paradox tells us that we need to be careful about inferring from compositional data. The conclusions made from a population may not be true for subpopulations and vice versa. 

\begin{table}[htbp]
  \centering
  \caption{An example of Simpson's paradox.}\label{tbl:simpson}
  \begin{subtable}{0.48\textwidth}
  \centering
   \caption{Number of students arriving on time and being late, classified by classroom and gender.} \label{tbl:simpson1}%
  \begin{tabular}{rlrr}
  \toprule
Classroom & Gender & On Time & Late \\
\midrule
      1     & M     & 53    & 9 \\
      1     & F     & 20    & 2 \\
      2     & M     & 12    & 6 \\
      2     & F     & 50    & 18 \\
      \bottomrule
      \end{tabular}%
  \end{subtable} %  
  \hfill
  \begin{subtable}{0.48\textwidth}
  \centering
\caption{Proportion of students arriving on time or being late, classified by classroom and gender.} \label{tbl:simpson2}%
\begin{tabular}{rlrr}
\toprule
Classroom & Gender & On Time & Late \\
\midrule
1     & M     &    0.855  &   0.145  \\
1     & F     &    \textbf{0.909}  &   0.091  \\
2     & M     &    0.667  &   0.333  \\
2     & F     &    \textbf{0.735}  &   0.265  \\
\bottomrule
\end{tabular}%
  \end{subtable}  
 
 \medskip
  
  \begin{subtable}{0.48\textwidth}
  \centering
\caption{Number of students arriving on time or being late, classified by gender.} \label{tbl:simpson3}%
\begin{tabular}{lrr}
\toprule
Gender & On Time & Late \\
\midrule
M     & 65    & 15 \\
F     & 70    & 20 \\
\bottomrule
\end{tabular}%
  \end{subtable} %  
  \hfill
  \begin{subtable}{0.48\textwidth}
  \centering
 \caption{Proportion of students arriving on time or being late, classified by gender.} \label{tbl:simpson4}%
  \begin{tabular}{lrr}
  \toprule
  Gender & On Time & Late \\
  \midrule
M     &    \textbf{0.813} &    0.188  \\
F     &    0.778  &    0.222  \\
\bottomrule
\end{tabular}%
    \end{subtable}
\end{table}%

Our motivation for studying compositional data regression in insurance comes from the facts that the recent explosion of telematics data analysis in insurance  \citep{Verbelen_2018,Guillen_2019,Pesantez_Narvaez_2019,Denuit_2019,Guillen_2020,So_2021} and that compositional data analysis has received little attention in the actuarial literature. In telematics data, we usually see total mileages traveled by an insured during the year and mileages traveled in different conditions such as at night, over speed, or in urban areas. The distances traveled in different conditions are examples of compositional data. Compositional data analysis has also been used to address ``coherence problem for subpopulations'' in forecasting mortality \citep{Bergeron_Boucher_2017}.

The classical method for dealing with compositional data is the logratio transformation method, which preserves the aforementioned properties. As pointed out by \cite{aitchison2003the}, logratio transformations cannot be applied to compositional data with zeros. In insurance data (e.g., telematics data), components of compositional vectors with zeros can be quite common. This sparsity imposes additional challenges to analyze compositional data in insurance.

In this paper, we investigate regression modeling of compositional variables (i.e., covariates) by using the exponential family principal component analysis (EPCA) to learn low dimensional representations of compositional data. Such representations of the compositional predictors extend the classical PCA as well as its probabilistic extension via probabilisitc principal components analysis (PPCA); both of which are special cases of EPCA. See \cite{Tipping_1999}.

The remaining part of the paper is organized as follows. In Section \ref{sec:review}, we review some literature related to compositional data analysis. In Section \ref{sec:data}, we describe the mine dataset used in our study. In Section \ref{sec:model}, we present three negative binomial regression models for fitting the data. In Section \ref{sec:result}, we provide numerical results of the proposed regression models. In Section \ref{sec:conclusion}, we conclude the paper with some remarks.

\section{Literature review}\label{sec:review}

Compositional data analysis dates back to the end of the 19th century when Karl Pearson published a paper about spurious correlations \citep{Pearson_1897}. Since then, the development of compositional data analysis has gone through four phases \citep{Aitchison_2005,pawlowsky2015modelling}.

The first phase started in 1897 when Karl Pearson introduced the concept of spurious correlation in his paper \citep{Pearson_1897} and ended around 1960. During this phase, standard multivariate statistical analysis was developed to analyze compositional data despite the fact that a compositional vector is subject to a constant-sum constraint. In particular, correlation analysis was used to analyze compositional vectors even though \cite{Pearson_1897} pointed out the pitfalls of interpreting correlations on proportions. 

The second phase started around 1960 when  the geologist Felix Chayes criticized the application of standard multivariate statistical analysis to compositional data \citep{Chayes_1960} and ended around 1980. \cite{Chayes_1960} criticized the interpretation of  correlation between components of a geochemical composition. During this phase, the main goal of compositional data analysis was to distort standard multivariate statistical techniques to analyze compositional data \citep{Aitchison_2005}.

The third phase started around 1980 and ended around 2000. In the early 1980s, the statistician John Aitchison realized that compositions provide only relative but not absolute information about values of components and that every statement about a composition can be stated in terms of ratios
of components \citep{Aitchison_1981,Aitchison_1982a,Aitchison_1983,Aitchison_1984}. During this phase, transformation techniques were developed to transform compositional data so that standard multivariate statistical analysis can be applied to the transformed data. In particular, a variety of logratio transformations were developed for compositional data analysis. The logratio transformation has the advantage that it provides a one-to-one mapping between compositional vectors, which stay in a constrained simplex space, and the associated logratio vectors, which stay in a unconstrained real space. In addition, any statement about compositions can be reformulated by logratios, and vice versa.

The fourth phase started around 2000 when researchers realized that the internal simplicial operation of perturbation, the external operation of powering, and the simplicial metric
define a metric vector space and that compositional data can be analyzed within this space with its specific algebraic-geometric structure. Research during this phase is characterized by developing staying-in-the-simplex approaches to
solve compositional problems. In a staying-in-the simplex approach, compositions are represented by orthonormal coordinates that live in a real Euclidean space.  The book by \cite{pawlowsky2015modelling} summarizes such approaches for compositional data analysis.

Researchers in the actuarial community have used compositional data in their studies, especially those related to telematics. In most existing actuarial literature, however, compositional data do not receive special treatment. For example, \cite{Guillen_2019} and \cite{Denuit_2019} selected some percentages of distances traveled in different conditions and treated them as normal explanatory variables. \cite{Pesantez_Narvaez_2019} treated compositional predictors as normal explanatory variables in their study of using XGBoost and logistic regression to predict motor insurance claims with telematics data. \cite{Guillen_2020} also treated compositional predictors as normal explanatory variables in their negative binomial regression models.

\cite{Verbelen_2018} used compositional data in generalized additive models and proposed a conditioning and a projection approach to handle structural zeros in components of compositional predictors. This is one of the few studies in insurance that consider handling compositional data. However, they did not consider dimension reduction techniques for compositional data in their regression models.

\section{Description of the data}\label{sec:data}

We use a dataset from the U.S. Mine Safety and Health Administration (MSHA) from 2013 to 2016. The dataset was used in the Predictive Analytics exam administrated by the Society of Actuaries in December 2018 \footnote{The dataset is available at \url{https://www.soa.org/globalassets/assets/files/edu/2018/2018-12-exam-pa-data-file.zip}.}. This dataset contains 53,746 observations described by 20 variables, including compositional variables. Table \ref{tbl:variable} shows the 20 variables and their descriptions. Among the 20 variables, 10 of them (i.e., the proportion of employee hours in different categories) are compositional components. 

\begin{table}[h!]
	\centering
	\caption{Description of the 20 variables of the mine dataset.}\label{tbl:variable}
	\begin{tabular}{lp{10cm}}
		\toprule
		\textbf{Variable} & \textbf{ Description }   \\
		\midrule
		\texttt{YEAR} & Calendar year of experience   \\
		\texttt{US\_STATE} & US state where mine is located    \\
		\texttt{COMMODITY} & Class of commodity mined   \\
		\texttt{PRIMARY} & Primary commodity mined    \\
		\texttt{SEAM\_HEIGHT} & Coal seam height in inches (coal mines only)   \\
		\texttt{TYPE\_OF\_MINE} & Type of mine    \\
		\texttt{MINE\_STATUS} & Status of operation of mine   \\
		\texttt{AVG\_EMP\_TOTAL} & Average number of employees    \\
		\texttt{EMP\_HRS\_TOTAL} & Total number of employee hours   \\
		\texttt{PCT\_HRS\_UNDERGROUND} & Proportion of employee hours in underground operations    \\
		\texttt{PCT\_HRS\_SURFACE} & Proportion of employee hours at surface operations of underground mine   \\
		\texttt{PCT\_HRS\_STRIP} & Proportion of employee hours at strip mine    \\
		\texttt{PCT\_HRS\_AUGER} & Proportion of employee hours in auger mining   \\
		\texttt{PCT\_HRS\_CULM\_BANK} & Proportion of employee hours in culm bank operations    \\
		\texttt{PCT\_HRS\_DREDGE} & Proportion of employee hours in dredge operations   \\
		\texttt{PCT\_HRS\_OTHER\_SURFACE} & Proportion of employee hours in other surface mining operations    \\
		\texttt{PCT\_HRS\_SHOP\_YARD } & Proportion of employee hours in independent shops and yards   \\
		\texttt{PCT\_HRS\_MILL\_PREP} & Proportion of employee hours in mills or prep plants    \\
		\texttt{PCT\_HRS\_OFFICE} & Proportion of employee hours in offices   \\
		\texttt{NUM\_INJURIES} & Total number of accidents reported   \\
		\bottomrule
	\end{tabular}%
	\label{tab:addlabel}%
\end{table}%

In our study, we ignore the following variables: \texttt{YEAR}, \texttt{US\_STATE}, \texttt{COMMODITY}, \texttt{PRIMARY}, \texttt{SEAM\_HEIGHT}, \texttt{MINE\_STATUS}, and \texttt{EMP\_HRS\_TOTAL}. We do not use \texttt{YEAR} in our regression models as we do not study the temporal effect of the data. However, we use \texttt{YEAR} to split the data into a training set and a test set. The variables \texttt{COMMODITY}, \texttt{PRIMARY}, and \texttt{SEAM\_HEIGHT} are related to the variable \texttt{TYPE\_OF\_MINE}, which we will use in our models. The variable \texttt{EMP\_HRS\_TOTAL} is highly correlated to the variable \texttt{AVG\_EMP\_TOTAL} with a correlation coefficient of 0.9942. We use only one of them in our models. 

\begin{table}[h!]
	\centering
	\caption{Summary of categorical variables.}\label{tbl:cat}
	\begin{tabular}{llll}
		\toprule
		YEAR  & Number of observations & TYPE\_OF\_MINE & Number of observations \\
		\midrule
		2013  & 13759 & Mill  & 2578 \\
		2014  & 13604 & Sand \& gravel & 25414 \\
		2015  & 13294 & Surface & 23091 \\
		2016  & 13089 & Underground & 2663 \\
		\bottomrule
	\end{tabular}%
\end{table}%

\begin{table}[htbp]
	\centering
	\caption{Summary of numerical variables.}\label{tbl:num}
	\begin{tabular}{lrrrrrr}
		\toprule
		Variable & {Min} & {1st Q} & {Median} & {Mean} & {3rd Q} & {Max} \\
		\midrule
		AVG\_EMP\_TOTAL  & 1     & 3     & 5     & 17.8419 & 12    & 3115 \\
		PCT\_HRS\_UNDERGROUND  & 0     & 0     & 0     & 0.0345 & 0     & 1 \\
		PCT\_HRS\_SURFACE  & 0     & 0     & 0     & 0.0087 & 0     & 1 \\
		PCT\_HRS\_STRIP  & 0     & 0.3299 & 0.8887 & 0.6801 & 1     & 1 \\
		PCT\_HRS\_AUGER  & 0     & 0     & 0     & 0.0046 & 0     & 1 \\
		PCT\_HRS\_CULM\_BANK  & 0     & 0     & 0     & 0.0046 & 0     & 1 \\
		PCT\_HRS\_DREDGE  & 0     & 0     & 0     & 0.045 & 0     & 1 \\
		PCT\_HRS\_OTHER\_SURFACE  & 0     & 0     & 0     & 0.0007 & 0     & 1 \\
		PCT\_HRS\_SHOP\_YARD  & 0     & 0     & 0     & 0.0036 & 0     & 1 \\
		PCT\_HRS\_MILL\_PREP  & 0     & 0     & 0     & 0.106 & 0     & 1 \\
		PCT\_HRS\_OFFICE  & 0     & 0     & 0.0312 & 0.1122 & 0.1685 & 1 \\
		NUM\_INJURIES  & 0 &0 &0 &0.4705 &0 &86 \\		
		\bottomrule
	\end{tabular}%
\end{table}%

Table \ref{tbl:cat} shows summary statistics of the categorical variable \texttt{TYPE\_OF\_MINE}. We also show the number of observations in different years. From the table, we see that there are approximately same number of observations from different years and that most mines are sand \& gravel and surface mines. 

\begin{table}[htb]
	\centering
	\caption{Percentage of zeros of the compositional variables.}\label{tbl:zero}
	\begin{tabular}{lr}
		\toprule
		Variable & {Percentage of zeros} \\
		\midrule
		\texttt{PCT\_HRS\_UNDERGROUND} & 95.34\% \\
		\texttt{PCT\_HRS\_SURFACE} & 95.99\% \\
		\texttt{PCT\_HRS\_STRIP} & 18.42\% \\
		\texttt{PCT\_HRS\_AUGER} & 99.49\% \\
		\texttt{PCT\_HRS\_CULM\_BANK} & 99.46\% \\
		\texttt{PCT\_HRS\_DREDGE} & 94.61\% \\
		\texttt{PCT\_HRS\_OTHER\_SURFACE} & 99.91\% \\
		\texttt{PCT\_HRS\_SHOP\_YARD } & 99.59\% \\
		\texttt{PCT\_HRS\_MILL\_PREP} & 81.24\% \\
		\texttt{PCT\_HRS\_OFFICE} & 36.93\% \\
		\bottomrule
	\end{tabular}%	
\end{table}

Table \ref{tbl:num} shows some summary statistics of the numerical variables. From the table, we see that the average number of employee ranges from 1 to 3115. All compositional components contain zeros. In additional, most compositional components are skewed as the 3rd quantiles are zero. Table \ref{tbl:zero} shows the percentages of zeros of the compositional variables. From the table, we see that most values of many compositional variables are just zeros.

\begin{figure}[htbp]
	\centering
	\begin{subfigure}{0.5\textwidth}
		\includegraphics[width=0.995\textwidth]{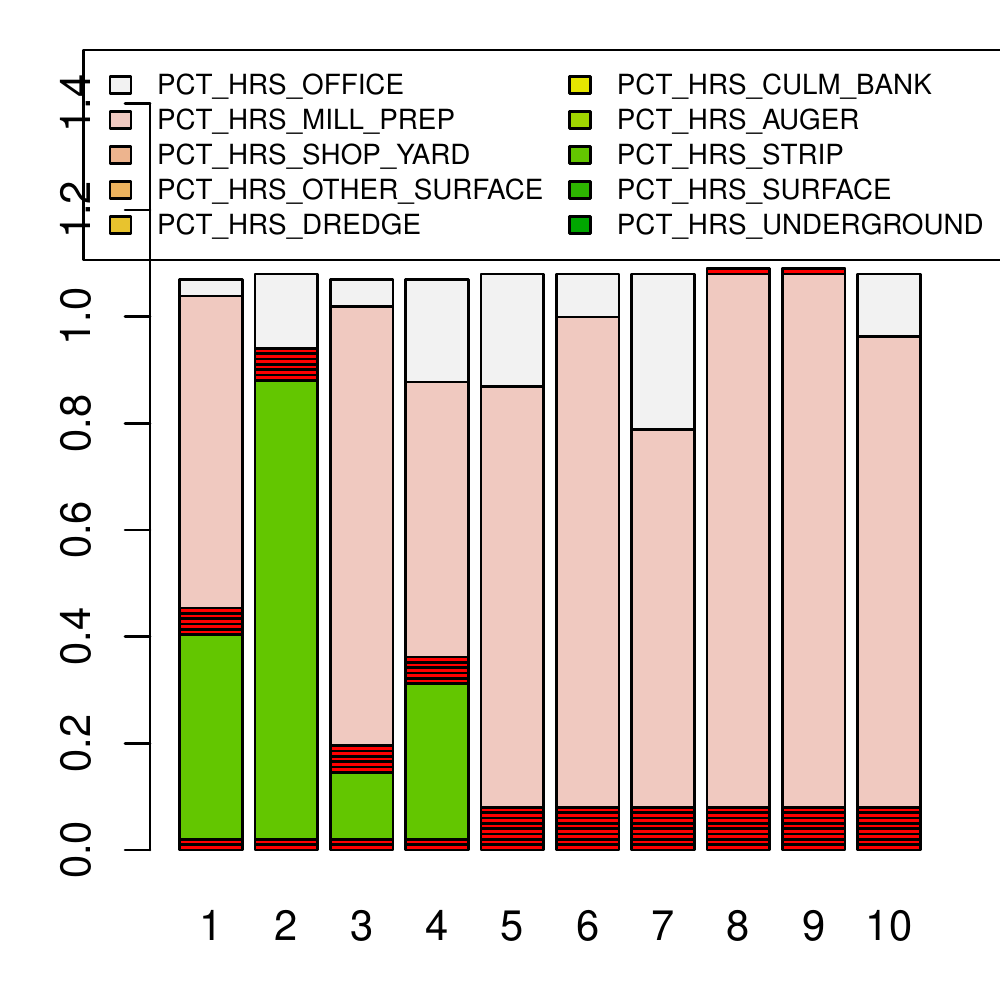}
		\caption{}\label{fig:mine}
	\end{subfigure}%
	\begin{subfigure}{0.5\textwidth}
		\includegraphics[width=0.995\textwidth]{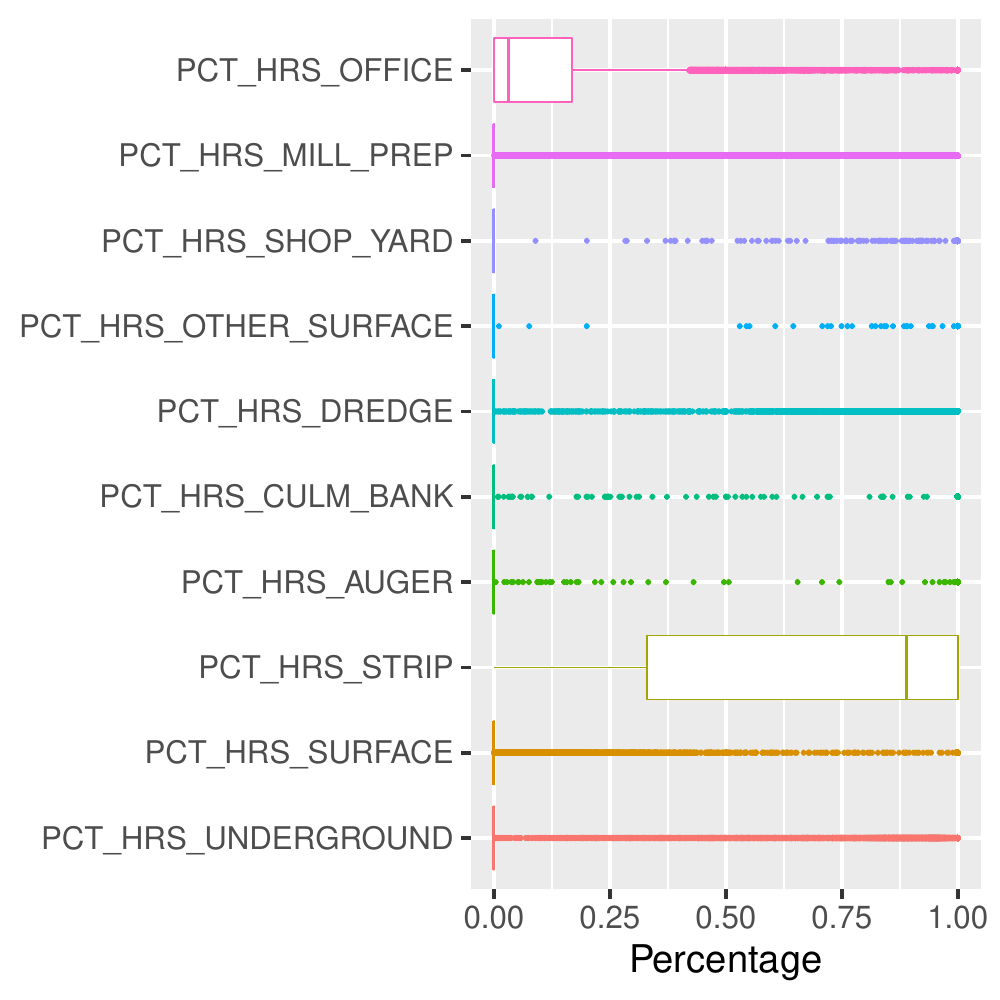}
		\caption{}\label{fig:pct}
	\end{subfigure}
	\caption{Compositional variables of the mine dataset. The left figure shows the barplots of the compositional components of 10 observations. The right figure shows the boxplots of the compositional covariates.}\label{fig:minepc}
\end{figure}

Figure \ref{fig:mine} shows the proportions of employee hours in different categories from the first ten observations. Figure \ref{fig:pct} shows the boxplots of the compositional variables. From the figures, we also see that the compositional variables have many zero values. Zeros in compositional data are irregular values. \cite{pawlowsky2015modelling} discussed several strategies to handle such irregular values. One strategy is to replace zeros by small positive values. In the mine dataset, we can treat zeros as rounded zeros below a detection limit. In this paper, we replace zero components with $10^{-6}$, which corresponds to a detection limit of one hour among one million employee hours.

\begin{figure}[htbp]
	\centering
	\includegraphics[width=0.8\textwidth]{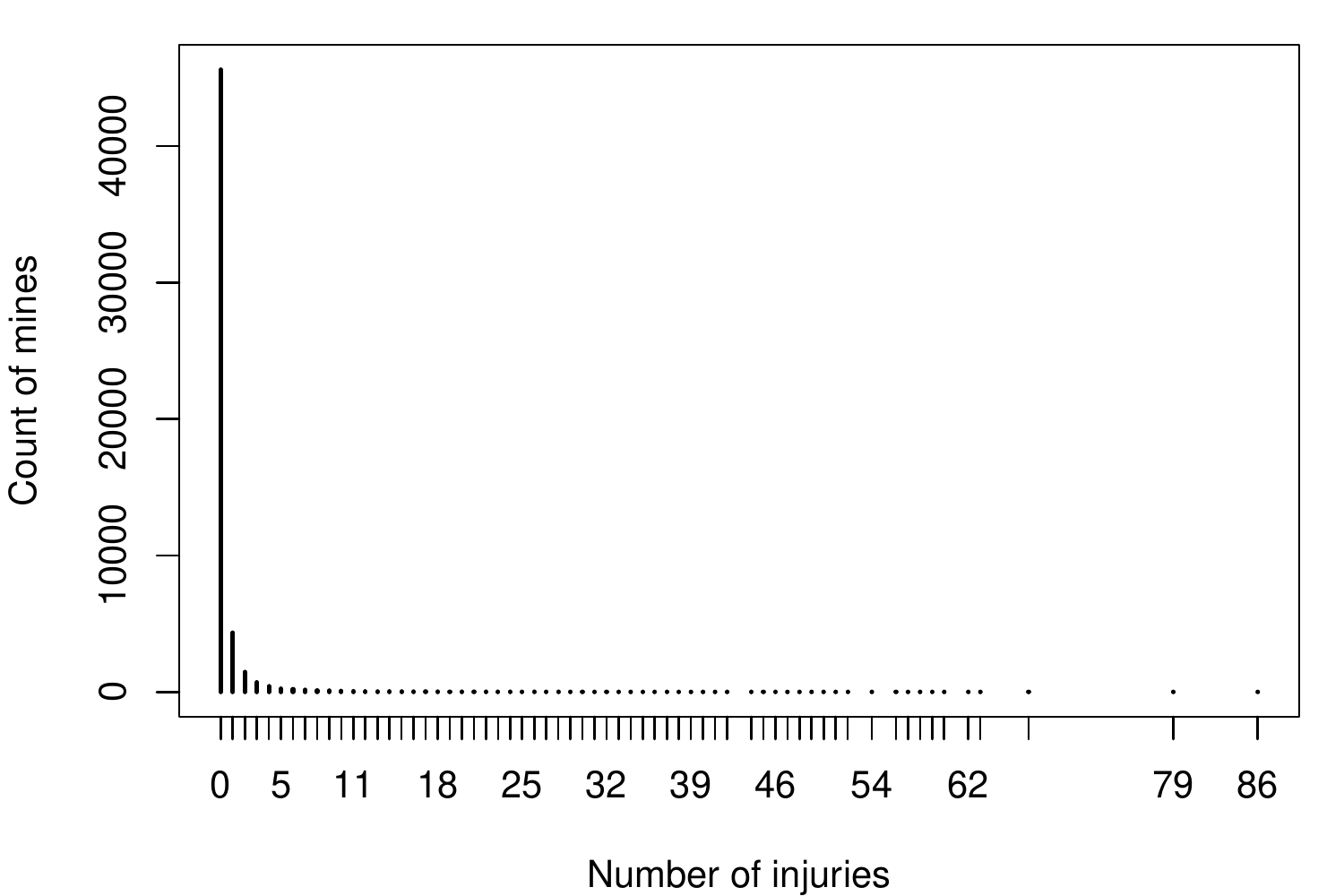}
	\caption{A frequency distribution of the response variable. }\label{fig:count}
\end{figure}

\begin{figure}[htbp]
	\centering
	\includegraphics[width=0.8\textwidth]{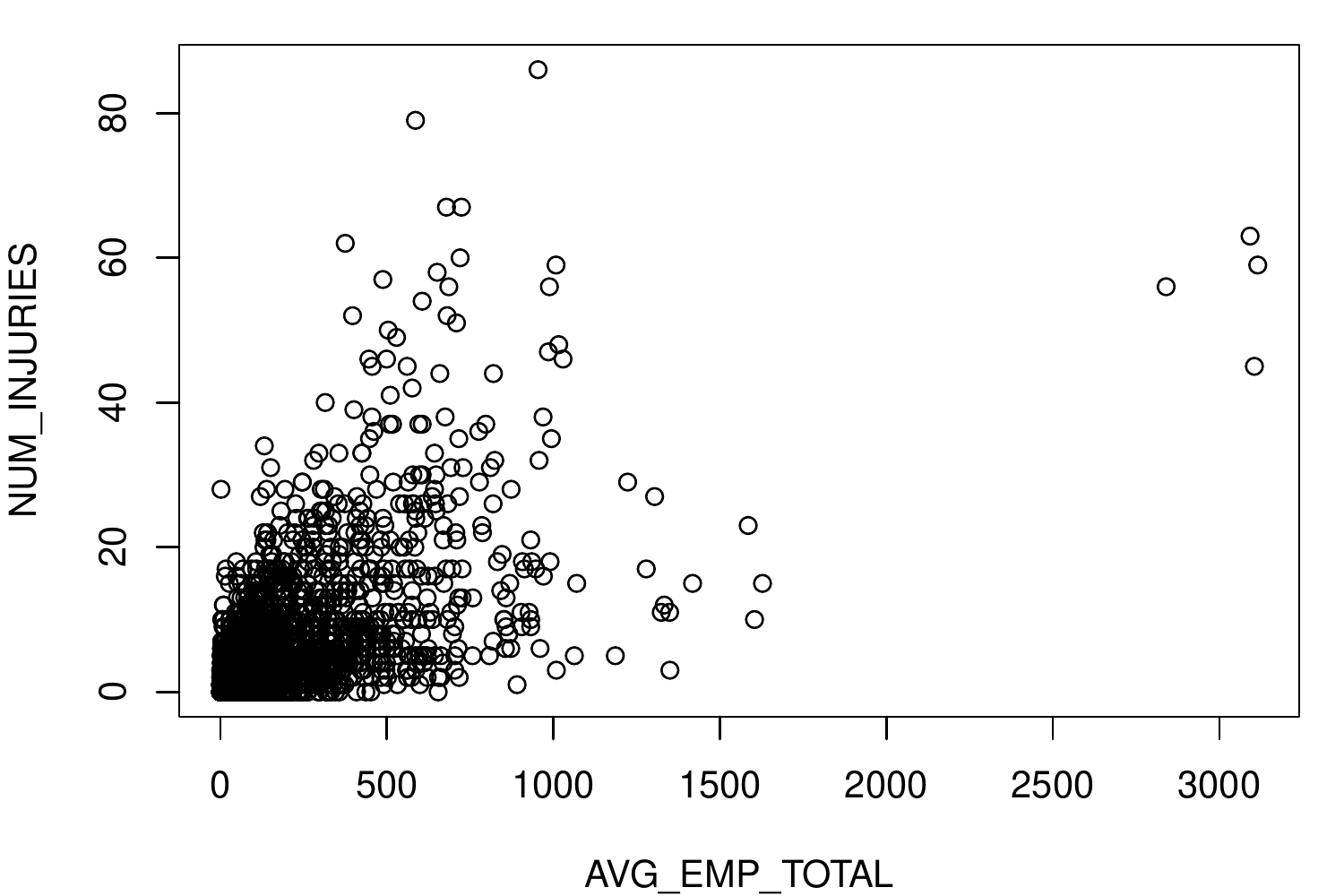}
	\caption{A scatter plot between the average number of employees and the total number of injuries reported. }\label{fig:exposure}
\end{figure}

The response variable is the total number of injuries reported. Figure \ref{fig:count} shows a frequency distribution of the response variable. The summary statistics given in Table \ref{tbl:num} and Figure \ref{fig:count} show that the response variable is also highly skewed with many zeros. Figure \ref{fig:exposure} shows a scatter plot between the variable \texttt{AVG\_EMP\_TOTAL} and the response variable \texttt{NUM\_INJURIES}. From the figure, we see that the two variables have a positive relationship. In our models, we will use  the variable \texttt{AVG\_EMP\_TOTAL} as the exposure.

\section{Models}\label{sec:model}

In this section, we present the models for the compositional data described in the previous section. In particular, we first present the logratio transformation in detail. Then we present a dimension reduction technique for compositional data. Finally, we present regression models with compositional covariates.

\subsection{Transformation}

The classical method for dealing with compositional data is the logratio transformation method, which preserves the aforementioned properties. Let $X=\{\bfx{x}_1,\ldots,\bfx{x}_n\}$ be a compositional dataset with $n$ compositional vectors. These vectors are assumed to be contained in the simplex:
\[
\mathbb{S}^d = \left\{\bfx{x}\in \mathbb{R}^d: \sum_{j=1}^d x_j = \kappa; \forall 
j, x_j > 0\right\},
\]
where $\kappa>0$ is a constant, usually 1. The superscript $d$ in $\mathbb{S}^d$ does not denote the dimensionality of the simplex as the dimensionality of the simplex is $d-1$.

The logratio transformation method transforms the data from the simplex to the real Euclidean space \citep{Aitchison_1994}. There are three types of logratio transformations: centered logratio transformation (CLR), additive logratio transformation (ALR), and isometric logratio transformation (ILR). The CLR transformation scales each compositional vector by its geometric mean; the ALR transformation applies log-ratio between a component and a reference
component; and the ILR transformation uses an orthogonal coordinate system in the simplex.

Let $\bfx{x}$ be a column vector (i.e., a $d\times 1$ vector). The CLR transformation is defined as:
\begin{equation}\label{eq:clr}
\clr(\bfx{x}) = \log\dfrac{\bfx{x}}{g(\bfx{x})}   = \log(\bfx{x}) - \frac{1}{d}\sum_{j=1}^d \log(x_j) = M_{c} \log(\bfx{x}),
\end{equation}
where $g(\bfx{x})=\left(\prod_{j=1}^d x_j\right)^{1/d}$ is the geometric mean of $\bfx{x}$ and $M_{c}=I_d - \frac{1}{d} \bfx{1}_d\bfx{1}_d^T$ is a $d\times d$ matrix. Here $I_d$ is a $d\times d$ identity matrix and $\bfx{1}_d$ is a $d\times 1$ vector of ones, i.e.,
\[
M_c = \begin{pmatrix}
	1 & 0 & \cdots & 0\\
	0 & 1 & \cdots & 0\\
	\vdots & \vdots & \ddots & \vdots\\
	0 & 0 & \cdots & 1
\end{pmatrix} - \dfrac{1}{d} \begin{pmatrix}
	1 \\
	1 \\
	\vdots\\
	1 
\end{pmatrix}\begin{pmatrix}
1 & 1 & \cdots & 1
\end{pmatrix} = \begin{pmatrix}
	1 & 0 & \cdots & 0\\
	0 & 1 & \cdots & 0\\
	\vdots & \vdots & \ddots & \vdots\\
	0 & 0 & \cdots & 1
\end{pmatrix} - \dfrac{1}{d} \begin{pmatrix}
1 & 1 & \cdots & 1\\
1 & 1 & \cdots & 1\\
\vdots & \vdots & \ddots & \vdots\\
1 & 1 & \cdots & 1
\end{pmatrix}.
\]

The ALR transformation is defined similarly. Suppose that the $d$th component of compositional vectors is selected as the reference component. Then the ALR transformation is defined as:
\begin{equation}\label{eq:alr}
	\alr(\bfx{x}) = \log(\bfx{x}) - \log(x_d) = = M_{a} \log(\bfx{x}),
\end{equation}
where $x_d$ is the $d$th component of $\bfx{x}$ and $M_a$ is a $(d-1)\times d$ matrix given by
\[
M_a = \begin{pmatrix}
	I_{d-1} & -\bfx{1}_{d-1}
\end{pmatrix} = \begin{pmatrix}
1 & 0 & \cdots & 0 &- 1\\
0 & 1 & \cdots & 0 & -1\\
\vdots & \vdots & \ddots & \vdots& -1\\
0 & 0 & \cdots & 1 & -1
\end{pmatrix}.
\]

The ILR transformation involves an orthogonal coordinate system in the simplex and is defined as:
\begin{equation}\label{eq:ilr}
	\ilr(\bfx{x}) =  R\cdot \clr(\bfx{x}),
\end{equation}
where $R$ is a $(d-1)\times d$ matrix that satisfies the following conditions:
\[
RR^T = I_{d-1}, \quad R^TR = M_c =I_d - \frac{1}{d} \bfx{1}_d\bfx{1}_d^T.
\]
The matrix $R$ is called the contrast matrix \citep{pawlowsky2015modelling}. The CLR and ILR transformations have the following relationship:
\[
R^T\ilr(\bfx{x}) = R^TR\clr(\bfx{x}) = R^TRR^TR \log(\bfx{x})=R^TR\log(\bfx{x}) = \clr(\bfx{x}).
\]
Further, we have
\[
R \clr(\bfx{x}) = RR^T\ilr(\bfx{x}) = \ilr(\bfx{x}).
\]

All three logratio transformation methods are equivalent up to linear transformations. The ALR transformation does not preserve distances. The CLR method preserves distances but can lead to singular covariance matrices. The IRL methods avoids these drawbacks. However, it is challenging to interpret the resulting coordinates.

These logratio transformation methods provide one-to-one mappings between the real Euclidean space and the simplex and allow transforming back to the simplex from the Euclidean space without losing of information. The reverse operation of the CLR transformation, for example, is given by:
\begin{equation}\label{eq:clr2}
	\bfx{x}= \clr^{-1}(\bfx{z}) = \dfrac{\kappa \exp(\bfx{z})}{ \sum_{j=1}^{d} \exp(z_j)},
\end{equation}
where $\bfx{z}=\clr(\bfx{x})$ is the transformation of $\bfx{x}$ under the CLR method. The reverse operation of the ILR transformation is given by:
\begin{equation}\label{eq:ilr2}
	\bfx{x}= \clr^{-1}(R^T \ilr(\bfx{x})),
\end{equation}
where $\clr^{-1}$ is defined in Equation \eqref{eq:clr2}.

\subsection{PCA for compositional data}

In this subsection, we introduce some dimension reduction methods for compositional data. In particular, we introduce the traditional principal component analysis (PCA) and the exponential family PCA (EPCA) for compositional data in details. The EPCA methodology has recently been shown to transform compositional data to uncover the true structure \citep{Avalos_2018}.

The EPCA method is a generalized PCA method for distributions from the exponential family that was proposed by \cite{Collins_2002} based on the same ideas of generalized linear models. To describe the traditional PCA and the EPCA method, let us first introduce the $\varphi$-PCA method because both the traditional PCA and the EPCA methods can be derived from the $\varphi$-PCA method. 

We consider a dataset $X=(\mathbf{x}_1,\ldots,\mathbf{x}_n)$ consisting of $n$ numerical vectors, each of which consists of $d$ components. For notation purpose, we assume all vectors are column vectors and $X$ is a $d\times n$ matrix. 

Let $\varphi: \mathbb{R}^d\rightarrow \mathbb{R}$ be a convex and differentiable function. Then the $\varphi$-PCA aims to find matrices $A\in \mathbb{R}^{l\times n}$ and $V\in \mathbb{R}^{l\times d}$ with $VV^T=I_l$ that minimize the following loss function:
\begin{equation}\label{eq:loss}
	\mathcal{L}_{\varphi-PCA}(X;A,V) = \sum_{i=1}^n D_{\varphi}(\mathbf{x}_i, V^T\mathbf{a}_i),
\end{equation}
where  $\mathbf{a}_i$ is the $i$th column of $A$ and $D_{\varphi}$ denotes the Bregman divergence:
\begin{equation}\label{eq:bregman}
	D_{\varphi}(\bfx{x}, \bfx{y}) = \varphi(\bfx{x}) - \varphi(\bfx{y}) - (\bfx{x}-\bfx{y})^T [\bigtriangledown\varphi(\bfx{y})].
\end{equation}
Here $\bigtriangledown$ denotes the vector differential operator. A Bregman divergence can be thought of as a general distance. Table \ref{tbl:bregman} gives two examples of convex functions and the corresponding Bregman divergences.

\begin{table}[htbp]
	\centering
	\caption{Examples of some convex functions and the corresponding Bregman divergences.}\label{tbl:bregman}
	\begin{tabular}{ll}
		\toprule
		$\varphi(\bfx{x})$ &  $D_{\varphi}(\bfx{x}, \bfx{y})$ \\
		\midrule
		$\frac{1}{2}\Vert \bfx{x}\Vert^2$ &  $\frac{1}{2}\Vert \bfx{x} - \bfx{y}\Vert^2$\\[3mm]
		$\sum_{j=1}^d \exp(x_j)$  &  $\sum_{j=1}^d e^{y_j}( e^{x_j -y_j} - 1  - x_j + y_j)$ \\
		\bottomrule
	\end{tabular}
\end{table}

For a distribution in the exponential family, the conditional probability of a value $x$ given a parameter value $\theta$ has the following form:
\begin{equation}\label{eq:pxtheta}
	\log P(x\vert\theta) = \log P_0(x) + x\theta - G(\theta),
\end{equation}
where $P_0(x)$ is a function of $x$ only, $\theta$ is the natural parameter of the distribution, and $G(\theta)$ is a function of $\theta$ that ensures that $P(x\vert\theta)$ is a density function. The negative log-likelihood function of the exponential family can be expressed through a Bregman divergence. To see this, let $F$ be a function defined by
\[
F(h(\theta)) + G(\theta) = h(\theta)\theta,
\]
where $h(\theta) = G'(\theta)$ is the derivative of $G$. From the above definition, we can show that $f(\theta)=h^{-1}(\theta)$. Hence
\begin{align*}
	-\log P(x\vert\theta) &= -\log P_0(x) - x\theta + G(\theta) \\
	& =  -\log P_0(x) - x\theta + h(\theta)\theta - F(h(\theta)) \\
	& =  -\log P_0(x) - F(x)  + F(x) - F(h(\theta)) - (x - h(\theta))\theta  \\
	& =  -\log P_0(x) - F(x)  + F(x) - F(h(\theta)) - (x - h(\theta))f(h(\theta))  \\
	& = -\log P_0(x) - F(x) + \varphi_F(x, h(\theta)),
\end{align*}
which shows that the negative log-likelihood can be written as a Bregman distance plus two terms that are constant with respect to $\theta$. As a result, minimizing the negative log-likelihood function with respect to $\theta$ is equivalent to minimizing the Bregman divergence.

The EPCA method aims to find matrices $A\in \mathbb{R}^{l\times n}$ and $V\in \mathbb{R}^{l\times d}$ with $VV^T=I_l$ that minimize the following loss function:
\begin{equation}\label{eq:epca1}
	\mathcal{L}_{EPCA}(X;A,V) = -\log P(X\vert A,V) = -\sum_{i=1}^n\sum_{j=1}^d \log P(x_{ji} \vert \theta_{ji}),
\end{equation}
where 
\[
\theta_{ji} = \sum_{h=1}^l a_{hi} v_{hj}
\]
or
\[
\boldsymbol{\Theta} = (\theta_{ji})_{d\times n} = V^TA.
\]
In other words, the EPCA method tries to find a lower dimensional subspace of parameter space to approximate the original data. The matrix $V$ contains $l$ basis vectors of $\mathbb{R}^d$ and $\boldsymbol{\theta}_i=(\theta_{1i},\theta_{2i},\ldots,\theta_{di})^T$ is represented as a linear combination of the basis vectors. The traditional PCA is a special case of EPCA when the normal distribution is appropriate for the data. In this case, $\varphi(\bfx{x}) = \frac{1}{2}\Vert \bfx{x}\Vert^2$. 

To apply the traditional PCA to compositional data, we first need to transform them from the simplex into the real Euclidean space by using a logratio transformation method (e.g., the CLR method). When the compositional data is transformed by the CLR method, the traditional PCA has the following loss function:
\begin{equation}\label{eq:pca}
	\mathcal{L}_{PCA}(X;A,V) = \dfrac{1}{2}\Vert\clr(X) - V^TA\Vert^2,
\end{equation}
where $\clr(X) = (\clr(\bfx{x}_1), \clr(\bfx{x}_2), \cdots, \clr(\bfx{x}_n))$ is the CLR transformed data.

\cite{Avalos_2018} proposed the following  EPCA method for compositional data:
\begin{equation}\label{eq:epca}
	\mathcal{L}_{EPCA}(X;A,V) = D_{exp}(\clr(X), V^TA ),
\end{equation}
where $D_{exp}(\cdot,\cdot)$ is the Bregman divergence corresponding to the $\exp(\cdot)$ function. \cite{Avalos_2018} also proposed an optimization procedure to find $A$ and $V$ such that the loss function is minimized.

\subsection{Regression with compositional covariates}

Compositional data in regression models refer to a set of predictor variables that describe parts of some whole and are commonly presented as vectors of proportions, percentages, concentrations, or frequencies.

\cite{pawlowsky2015modelling} presented a linear regression model with compositional covariates. The linear regression model is formulated as follows:
\begin{equation}\label{eq:lm}
	\hat{y}_i = \beta_0 +\langle \boldsymbol{\beta}, \bfx{x}_i\rangle_a,
\end{equation}
where $\beta_0$ is a real intercept and $\langle\cdot,\cdot\rangle_a$ is the Aitchison inner product defined by
\[
\langle \boldsymbol{\beta}, \bfx{x}_i\rangle_a = \langle \clr(\boldsymbol{\beta}), \clr(\bfx{x}_i)\rangle=\langle \ilr(\boldsymbol{\beta}), \ilr(\bfx{x}_i)\rangle=\sum_{j=1}^d \log\dfrac{\beta_j}{g(\boldsymbol{\beta})}\log\dfrac{x_j}{g(\bfx{x})}.
\]
Here $g(\cdot)$ denotes the geometric mean. The sum of squared errors is given by:
\begin{equation}\label{eq:sse}
	SSE = \sum_{i=1}^n \left(y_i - \beta_0 - \langle \boldsymbol{\beta}, \bfx{x}_i\rangle_a \right)^2 = \sum_{i=1}^n \left(y_i - \beta_0 - \langle \ilr(\boldsymbol{\beta}), \ilr(\bfx{x}_i)\rangle \right)^2,
\end{equation}
which suggests that the actual fitting can be done using the IRL transformed coordinates, that is, a linear regression can be simply fitted to the response as a linear function of $\ilr(\bfx{x})$. The CLR transformation should not be used because it requires the generalized inversion of singular matrices.

Since the response variable of the mine dataset is a count variable, it is not suitable to use linear regression models. Instead, we will use generalized linear models to model the count variable. Common generalized linear models for count data include Poisson regression models and negative binomial models \citep{frees2009}. Poisson models are simple but restrictive as they assume that the mean and the variance of the response are equal. For the mine dataset, the mean and the variance of the response variable \texttt{NUM\_INJURIES} are 0.4705 and 5.4636, respectively.  In this study, we use negative binomial models as they provide more flexibility than Poisson models. 

Let $\bfx{x}_1$, $\bfx{x}_2$, $\cdots$, $\bfx{x}_n$ denote the predictors from $n$ observations. The predictors can be selected in different ways. For example, the predictors can be the IRL transformed data or the principle components produced by a dimension reduction method such as the traditional PCA and the EPCA methods. For $i=1,2,\ldots,n$, let $y_i$ be the response corresponding to $\bfx{x}_i$. In a negative binomial regression model, we assume that the response variable follows a negative binomial distribution:
\begin{equation}\label{eq:nbd}
	P(y=j) = \begin{pmatrix}
		j+r-1 \\
		r-1
	\end{pmatrix} p^r (1-p)^j,
\end{equation}
where $r>0$ and $p\in (0,1)$ are parameters.  The mean and the variance of a variable $y$ are given by
\[
E[y] = \frac{r(1-p)}{p},\quad \var{(y)} = \frac{r(1-p)}{p^2}.
\]

To incorporate covariates in a negative binomial regression model, we let the parameter $p$ to vary by observation. In particular, we incorporate the covariates as follows:
\begin{equation}\label{eq:nbr}
	\mu_i = \frac{r(1-p_i)}{p_i} = E_i \exp\left(\bfx{x}_i'\boldsymbol{\beta}\right),
\end{equation}
where $r$ and $\boldsymbol{\beta}$ are parameters to be estimated, and $E_i$ is the exposure. For the mine dataset, we use the variable \texttt{AVG\_EMP\_TOTAL} as the exposure. This is similar to model the number of claims in a Property \& Casualty dataset where the time to maturity of an auto insurance policy is used as the exposure \citep{frees2009,gan2018}. The method of maximum likelihood can be used to estimate the parameters.

\begin{table}[htbp]
	\centering
	\caption{Description of negative binomial regression models for the mine dataset.}\label{tbl:model}
	\begin{tabular}{ll}
		\toprule
		Model & Covariates \\
		\midrule
		NB & \texttt{TYPE\_OF\_MINE} and the IRL transformed compositional components \\
		NBPCA & \texttt{TYPE\_OF\_MINE} and principle components from the traditional PCA method\\
		NBEPCA & \texttt{TYPE\_OF\_MINE} and principle components from the EPCA method \\
		\bottomrule
	\end{tabular}
\end{table}

In this study, we compare three negative binomial models. Table \ref{tbl:model} describes the covariates of the three models. In the first model, we use the compositional components transformed by the IRL method that is defined in Equation \eqref{eq:ilr}. In the second model, we use the first few principle components obtained from the traditional PCA method that is defined in Equation \eqref{eq:pca}. In the third model, we use the first few principle components obtained from the EPCA method that is defined in Equation \eqref{eq:epca}.

\section{Results}\label{sec:result}

In this section, we present the results of fitting the three negative binomial regression models (see Table \ref{tbl:model}) to the mine dataset. We follow the following procedure to fit and test the models:
\begin{enumerate}
	\item Transform the compositional variables for the whole mine dataset. For the model NB, we transform the compositional variables with the ILR method. For the model NBPCA, we first transform the compositional variables with the CLR method and then apply the standard PCA to the transformed data. Both the ILR method and the CLR method are available in the R package \texttt{compositions}. For the model NBEPCA, we transform the compositional variable with the EPCA method implemented by \cite{Avalos_2018} \footnote{The R code of the EPCA method is available at \url{https://github.com/sistm/CoDa-PCA}.}.
	\item Split the transformed data into a training set and a test set. In particular, we use data from years 2013 to 2015 as the training data and use data from 2016 as the test data.
	\item Estimate parameters based on the training set.
	\item Make predictions for the test set and calculate the out-of-sample validation measures.
\end{enumerate}

\subsection{Validation measure}

To measure the accuracy of regression models for count variables, it is common to use the Pearson's chi-square statistic, which is defined as:
\begin{equation}\label{eq:pchi}
	\chi^2 = \sum_{j=0}^{m} \frac{ (O_j - E_j)^2 }{E_j},
\end{equation}
where $O_j$ is the observed number of mines that have $j$ injuries, $E_j$ is the predicted number of mines that have $j$ injuries, $m$ is the maximum number of injuries considered over all mines. Between two models, the model with a lower chi-square statistic is better.

The observed number of mines that have $j$ injuries is obtained by
\[
O_j = \sum_{i=1}^n I_{\{y_i=j\}}, \quad j=0,1,\ldots,m,
\]
where $I$ is an indicator function. The predicted number of mines that have $j$ injuries is calculated as follows:
\begin{align*}
	E_j &= E\left[ \sum_{i=1}^n I_{\{\hat{y}_i=j\}} \right] = \sum_{i=1}^n E[I_{\{\hat{y}_i=j\}}] = \sum_{i=1}^n P(\hat{y}_i=j) \\
	&= \begin{pmatrix}
		j+\hat{r}-1 \\
		\hat{r}-1
	\end{pmatrix} \hat{p}_i^{\hat{r}} (1-\hat{p}_i)^j,
\end{align*}
where $\hat{r}$ and $\hat{p}_i$ are estimated values of the parameters. 

From Figure \ref{fig:count}, we see that the maximum observed number of injuries is 86. To calculate the validation measure, we use $m=99$, that is, we use the first 100 terms of $O_j$s and $E_j$s. The remaining terms are too close to zero and will be ignored.

\subsection{Results}

We fitted the three models described in Table \ref{tbl:model} to the mine dataset according to the aforementioned procedure. Table \ref{tbl:measure} shows the in-sample and out-of-sample chi-square statistics produced by the three models. From the table, we see that the negative binomial model with EPCA transformed data performs the best among the three models. The model NBEPCA produced the lowest chi-square statistics for both the training data and the test data.

\begin{table}[htbp]
	\centering
	\caption{In-sample and out-of-sample validation measures produced by the three models.}	\label{tbl:measure}%
	\begin{tabular}{lrr}
		\toprule
		Model &{In-sample $\chi^2$} & {Out-of-sample $\chi^2$} \\
		\midrule
		NB    & 334.9093 & 113.1498 \\
		NBPCA & 338.4336 & 114.2693 \\
		NBEPCA & \textbf{303.9178} & \textbf{111.8998} \\
		\bottomrule
	\end{tabular}%
\end{table}%

\begin{figure}[htbp]
	\centering
	\begin{subfigure}{0.5\textwidth}
		\includegraphics[width=0.995\textwidth]{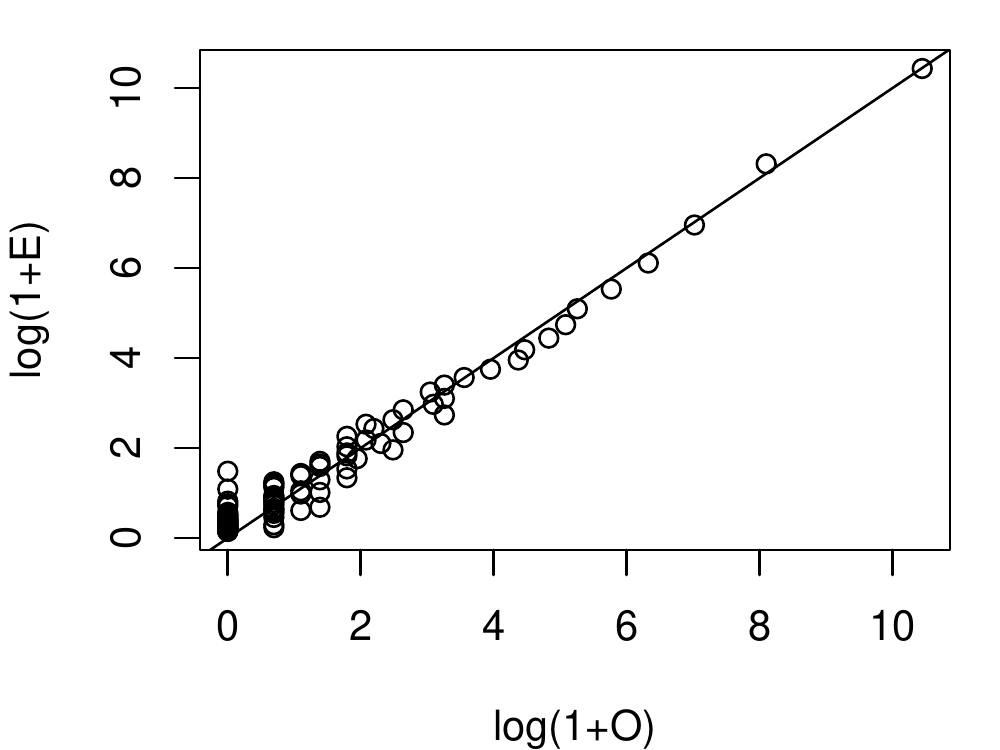}
		\caption{NB in-sample}\label{fig:res1a}
	\end{subfigure}%
	\begin{subfigure}{0.5\textwidth}
		\includegraphics[width=0.995\textwidth]{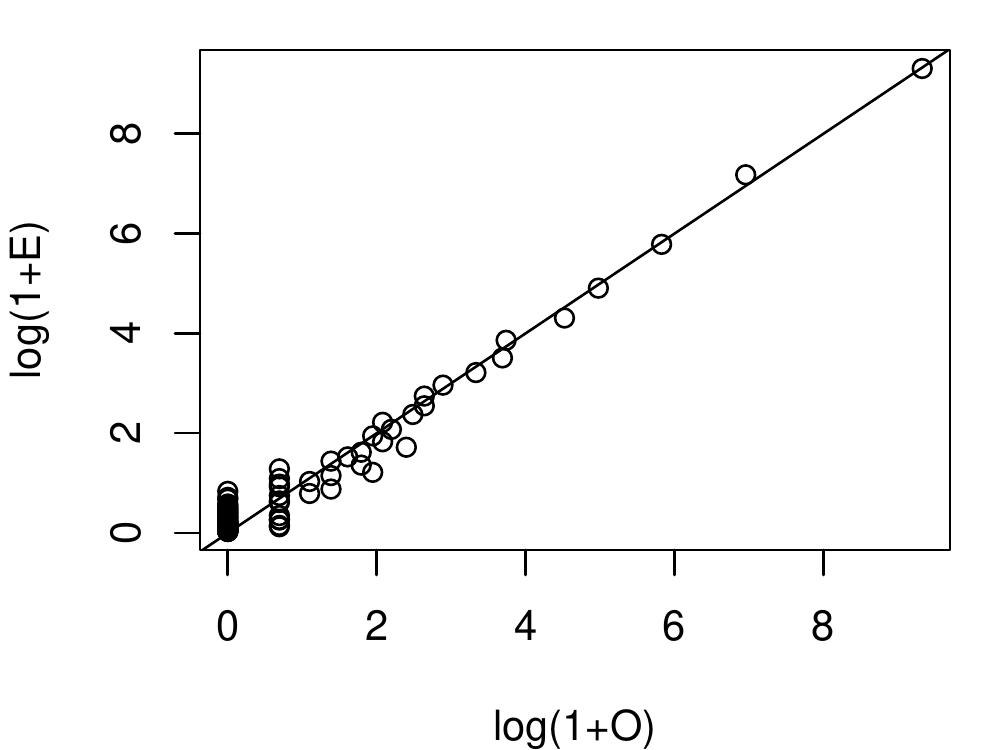}
		\caption{NB out-of-sample}\label{fig:res1b}
	\end{subfigure}
\begin{subfigure}{0.5\textwidth}
	\includegraphics[width=0.995\textwidth]{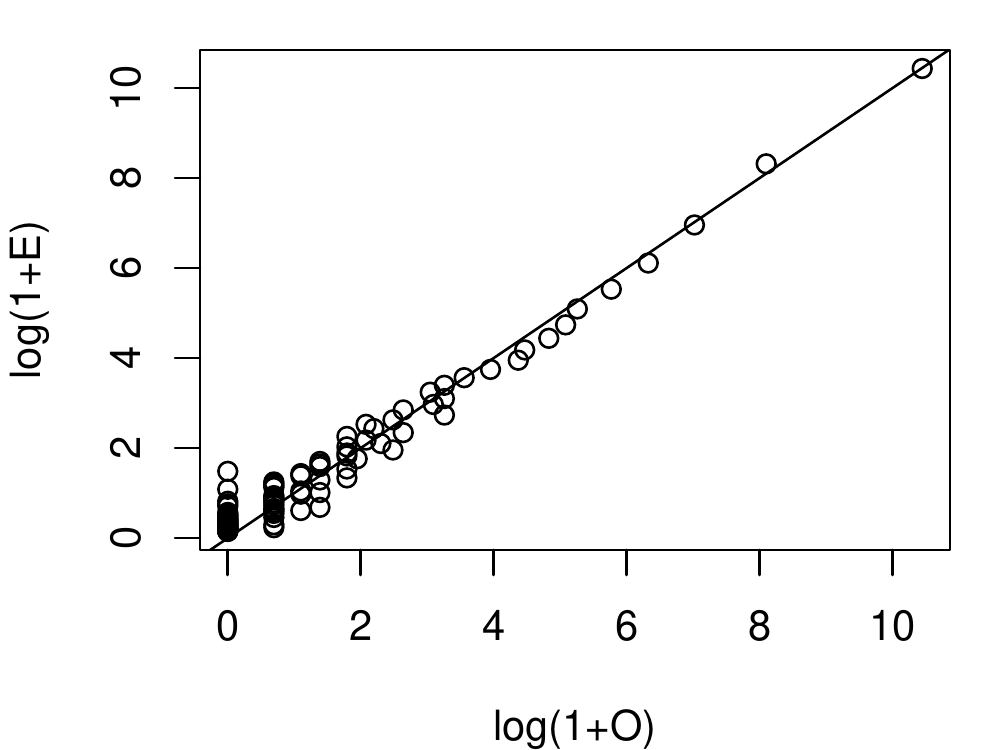}
	\caption{NBPCA in-sample}\label{fig:res2a}
\end{subfigure}%
\begin{subfigure}{0.5\textwidth}
	\includegraphics[width=0.995\textwidth]{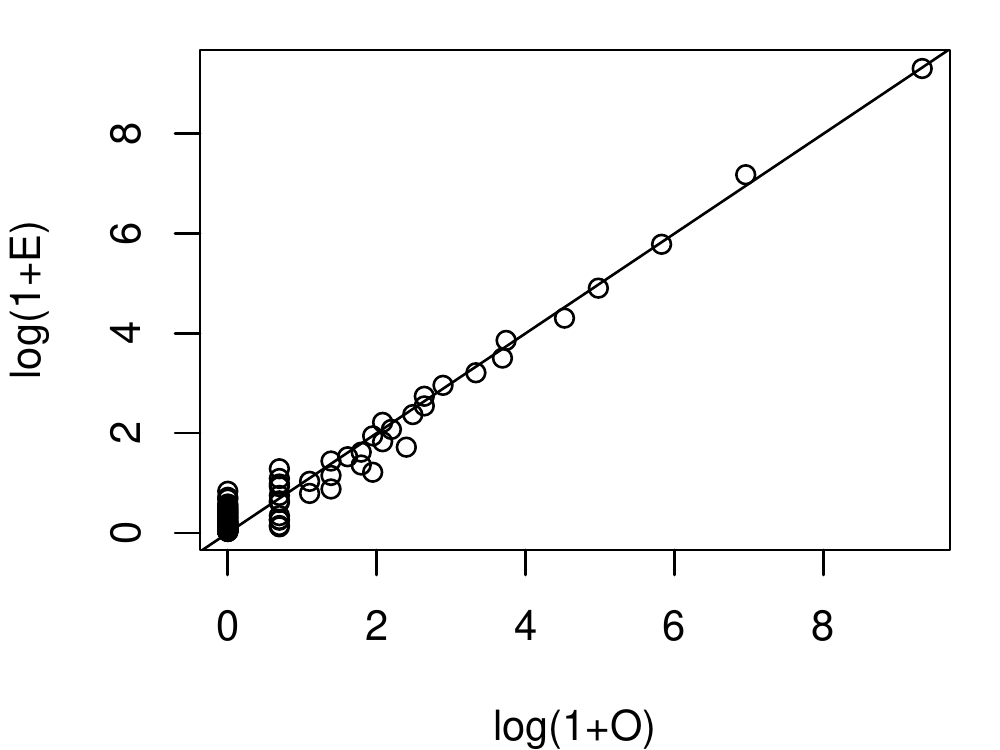}
	\caption{NBPCA out-of-sample}\label{fig:res2b}
\end{subfigure}
\begin{subfigure}{0.5\textwidth}
	\includegraphics[width=0.995\textwidth]{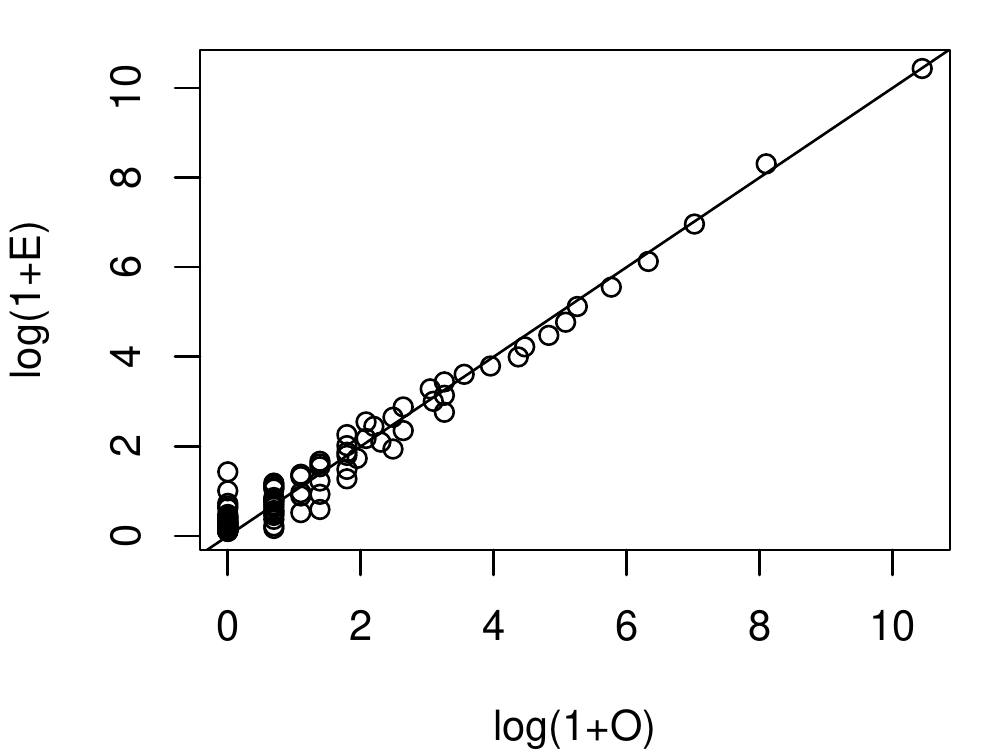}
	\caption{NBEPCA in-sample}\label{fig:res3a}
\end{subfigure}%
\begin{subfigure}{0.5\textwidth}
	\includegraphics[width=0.995\textwidth]{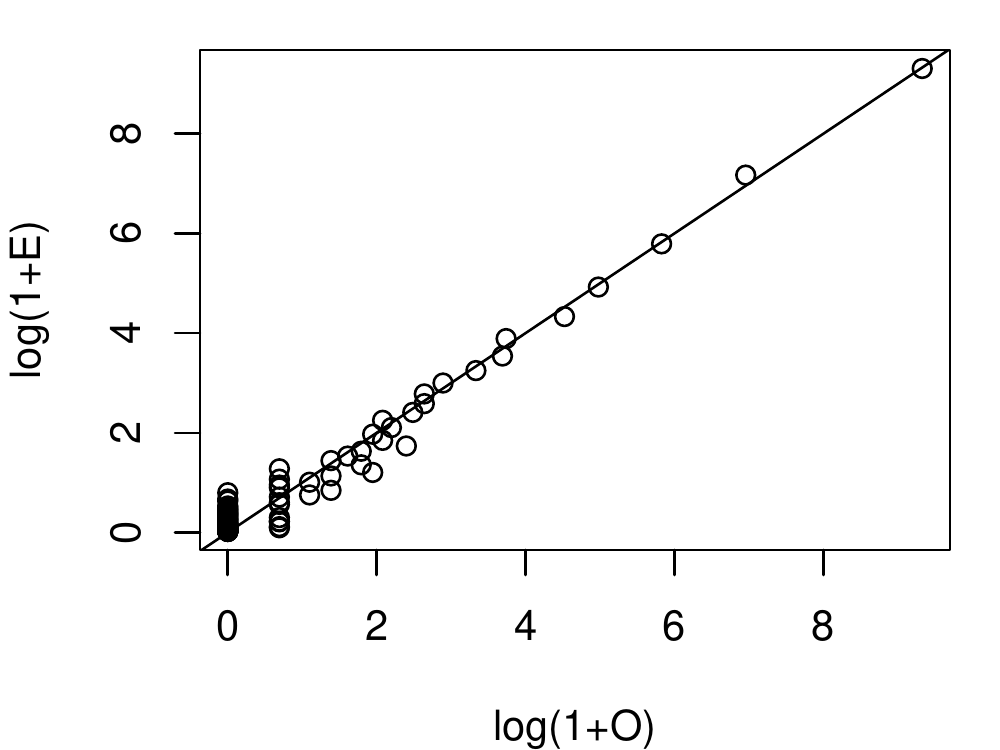}
	\caption{NBEPCA out-of-sample}\label{fig:res3b}
\end{subfigure}
	\caption{Scatter plots of the observed number of mines with different number of injuries and that predicted by models at log scales.}\label{fig:res}
\end{figure}

Figure \ref{fig:res} shows the scatter plots of the observed number of mines with different number of injuries and that predicted by models at log scales. That is, the scatter plots show the relationship between $\log(1+O_j)$ and $\log(1+E_j)$ for $j=0,1,\ldots,99$. We use log scale here because the $O_j$s and $E_j$s have big differences for different $j$s. From the scatter plots, we see that all the models predict the number of mines with different number of injuries quite well. We cannot tell the differences of the three models from the scatter plots. However, the chi-square statistics shown in Table \ref{tbl:measure} can tell that the NBEPCA model is the best.

\begin{figure}[htbp]
	\centering
	\includegraphics[width=0.8\textwidth]{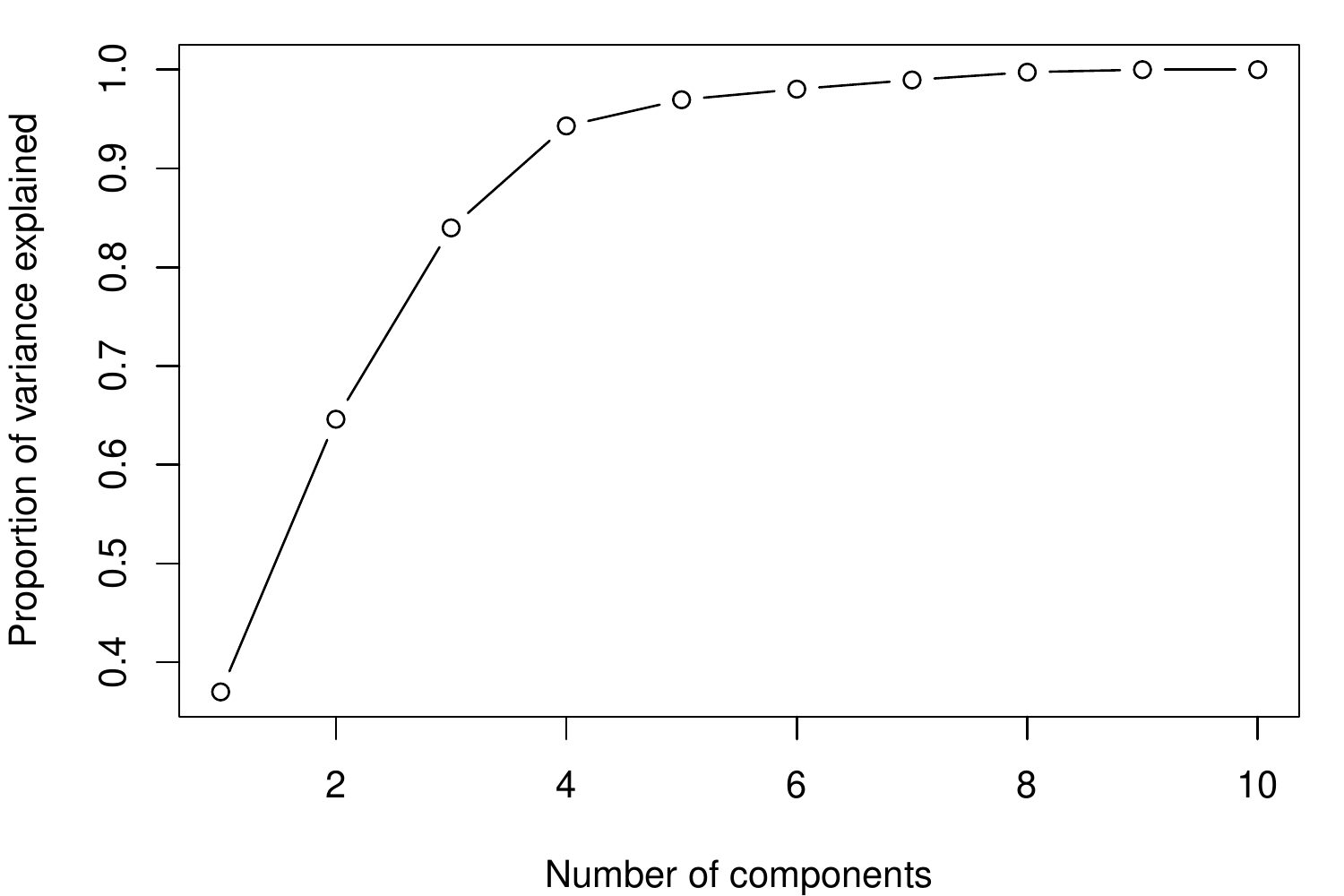}
	\caption{Proportion of variation explained with different principle components. }\label{fig:cumsum}
\end{figure}

For the NBPCA model, we selected the first five principal components produced by applying the standard PCA to the CLR transformed data. Figure \ref{fig:cumsum} shows the proportion of variation explained by different number of principal components. The first five principal components can explain more than 95\% of the variation of the original data. For the NBEPCA model, we also selected the first five principal components. The EPCA method used to transform the compositional variables is slow and took a normal desktop about 20 minutes to finish the computation. The reason is that the EPCA method we used is implemented in R, which is a scripting language and is not efficient for iterative optimization.

\begin{table}[htbp]
	\caption{Estimated regression coefficients of the model NB. Table (a) shows the coefficients of ILR transformed data. Table (b) shows the coefficients transformed back to the CLR proportions. }\label{tbl:coeff1}
	\begin{subtable}{\textwidth}
	\centering
	\caption{}\label{tbl:coeff1a}
\begin{tabular}{lrrrr}
	\toprule
	Predictor & {Coefficient} & {Std. Error} & {z value} & {P value} \\
	\midrule
	(Intercept)  & -3.7525 & 0.0505 & -74.3196 & 0 \\
	Sand \& gravel  & -0.2356 & 0.0795 & -2.9651 & 0.003 \\
	Surface  & 0.0444 & 0.0689 & 0.6446 & 0.5192 \\
	Underground  & -0.7147 & 0.1614 & -4.4288 & 0 \\
	V1    & -0.0536 & 0.0111 & -4.8275 & 0 \\
	V2    & -0.053 & 0.0061 & -8.6443 & 0 \\
	V3    & -0.0248 & 0.0127 & -1.9471 & 0.0515 \\
	V4    & -0.0674 & 0.0168 & -4.0039 & 0.0001 \\
	V5    & 0.0003 & 0.007 & 0.0437 & 0.9652 \\
	V6    & -0.044 & 0.0224 & -1.9631 & 0.0496 \\
	V7    & -0.018 & 0.013 & -1.3886 & 0.1649 \\
	V8    & 0.0143 & 0.0025 & 5.6681 & 0 \\
	V9    & -0.0057 & 0.0028 & -2.0133 & 0.0441 \\
		\bottomrule
	\end{tabular}%	
	\end{subtable}
\vskip 1em
\begin{subtable}{\textwidth}
	\centering
	\caption{}\label{tbl:coeff1b}
	\begin{tabular}{lr}
		\toprule
		Predictor & {Coefficient} \\
		\midrule
		PCT\_HRS\_UNDERGROUND  & 0.1093 \\
		PCT\_HRS\_SURFACE  & 0.1013 \\
		PCT\_HRS\_STRIP  & 0.0986 \\
		PCT\_HRS\_AUGER  & 0.1001 \\
		PCT\_HRS\_CULM\_BANK  & 0.0949 \\
		PCT\_HRS\_DREDGE  & 0.1008 \\
		PCT\_HRS\_OTHER\_SURFACE  & 0.0961 \\
		PCT\_HRS\_SHOP\_YARD  & 0.0982 \\
		PCT\_HRS\_MILL\_PREP  & 0.1013 \\
		PCT\_HRS\_OFFICE  & 0.0994 \\
		\bottomrule
	\end{tabular}%	
\end{subtable}
\end{table}

\begin{table}[h!]
	\caption{Estimated regression coefficients of the models NBPCA and NBEPCA.}\label{tbl:coeff23}
	\begin{subtable}{\textwidth}
		\centering
		\caption{NBPCA}\label{tbl:coeff2}
\begin{tabular}{lrrrr}
	\toprule
	Predictor & {Coefficient} & {Std. Error} & {z value} & {P value} \\
	\midrule
	(Intercept)  & -3.8635 & 0.0655 & -58.9829 & 0 \\
	Sand \& gravel  & -0.2261 & 0.0789 & -2.8656 & 0.0042 \\
	Surface  & 0.0588 & 0.0681 & 0.8631 & 0.3881 \\
	Underground  & -0.3017 & 0.1319 & -2.2876 & 0.0222 \\
	PC1   & 0.0274 & 0.0033 & 8.3922 & 0 \\
	PC2   & 0.0245 & 0.0035 & 7.0887 & 0 \\
	PC3   & -0.0007 & 0.0031 & -0.2257 & 0.8214 \\
	PC4   & -0.0356 & 0.0071 & -4.9989 & 0 \\
	PC5   & -0.0514 & 0.0096 & -5.3658 & 0 \\
			\bottomrule
		\end{tabular}%	
	\end{subtable}
\vskip 1em
	\begin{subtable}{\textwidth}
		\centering
		\caption{NBEPCA}\label{tbl:coeff3}
\begin{tabular}{lrrrr}
	\toprule
	Predictor & {Coefficient} & {Std. Error} & {z value} & {P value} \\
	\midrule
	(Intercept)  & -3.2533 & 0.1209 & -26.9052 & 0 \\
	Sand \& gravel  & -0.348 & 0.0719 & -4.8394 & 0 \\
	Surface  & -0.1203 & 0.0618 & -1.9444 & 0.0518 \\
	Underground  & -0.1894 & 0.1153 & -1.6428 & 0.1004 \\
	PC1   & 0.156 & 0.0184 & 8.479 & 0 \\
	PC2   & 0.0943 & 0.0073 & 12.8948 & 0 \\
	PC3   & -0.0414 & 0.0124 & -3.329 & 0.0009 \\
	PC4   & 0.0314 & 0.0152 & 2.0676 & 0.0387 \\
	PC5   & -0.2678 & 0.027 & -9.9187 & 0 \\
			\bottomrule
		\end{tabular}%	
	\end{subtable}
\end{table}

Table \ref{tbl:coeff1} shows the estimated regression coefficients of the model NB. As shown in Table \ref{tbl:coeff1a}, the ten compositional variables are transformed to nine variables by the ILR method. Table \ref{tbl:coeff1b} shows the regression coefficients transformed back to the CLR proportions by the inverse operation of the ILR method. From Table \ref{tbl:coeff1b}, we see that the coefficients of the ten compositional variables are similar and are quite different from the coefficients of the ILR variables $V1$ to $V9$. From the $P$ values shown in Table \ref{tbl:coeff1a}, we see that some of the IRL transformed variables are not significant. For example, the variable V5 has a $P$ value of 0.9652 and the variable V7 has a $P$ value of 0.1649. Both variables have a $P$ value greater than 0.1. This suggests that it is appropriate to reduce the dimensionality of the compositional variables.

The regression coefficients of the ILR transformed variables $V1$ to $V9$ shown in Table \ref{tbl:coeff1} vary a lot. It is challenging to interpret those coefficients. From the $P$ values shown in Table \ref{tbl:coeff1a}, we see that some of the IRL transformed variables are not significant. For example, the variable V5 has a $P$ value of 0.9652 and the variable V7 has a $P$ value of 0.1649. Both variables have a $P$ value greater than 0.1. This suggests that it is appropriate to reduce the dimensionality of the compositional variables.

Table \ref{tbl:coeff23} shows the estimated regression coefficients of the models NBPCA and NBEPCA. We used five principal components in both models. From Table \ref{tbl:coeff3}, we see that all principal components produced by the EPCA method are significant. However, Table \ref{tbl:coeff2} shows that the third principal component PC3 produced by the traditional PCA method is not significant as it has a $P$ value of 0.8214. This again shows the EPCA method is better than the traditional PCA method for the mine dataset.

In summary, our numerical results show that the EPCA method is able to produce better principal components than the traditional PCA method and that using the principal components produced by the EPCA method can improve the prediction accuracy of the negative binomial models with compositional covariates.

\section{Conclusions}\label{sec:conclusion}

Compositional data are commonly presented as vectors of proportions, percentages, concentrations, or frequencies. A peculiarity of these vectors is that their sum is constrained to be some fixed constant, e.g. 100\%.  Due to such constraints, compositional data have the following properties: they carry only relative information (scale invariance); equivalent results should be yielded when the ordering of the parts in the composition is changed (permutation invariance); and  results should not change if a noninformative part is removed (subcomposition coherence). 

In this paper, we investigated regression models with compositional covariates by using a mine dataset. We built negative binomial regression models to predict the number of injuries from different mines. In particular, we built three negative binomial regression models: a model with ILR transformed compositional variables, a model with principal components produced by the traditional PCA, and a model with principal components produced by the exponential family PCA. Our numerical results show that the exponential family PCA is able to produce principal components that are significant predictors and improve the prediction accuracy of the regression model.
 
\section*{Acknowledgments}

Guojun Gan and Emiliano A. Valdez would like to acknowledge the financial support provided by the Committee on Knowledge Extension Research of the Casualty Actuarial Society (CAS). 

\bibliographystyle{apalike}
\bibliography{pccomp}

\begin{thebibliography}{}

\bibitem[Aitchison, 1981]{Aitchison_1981}
Aitchison, J. (1981).
\newblock A new approach to null correlations of proportions.
\newblock {\em Journal of the International Association for Mathematical
  Geology}, 13(2):175--189.

\bibitem[Aitchison, 1982]{Aitchison_1982a}
Aitchison, J. (1982).
\newblock The statistical analysis of compositional data.
\newblock {\em Journal of the Royal Statistical Society: Series B
  (Methodological)}, 44(2):139--160.

\bibitem[Aitchison, 1983]{Aitchison_1983}
Aitchison, J. (1983).
\newblock Principal component analysis of compositional data.
\newblock {\em Biometrika}, 70(1):57--65.

\bibitem[Aitchison, 1984]{Aitchison_1984}
Aitchison, J. (1984).
\newblock The statistical analysis of geochemical compositions.
\newblock {\em Journal of the International Association for Mathematical
  Geology}, 16(6):531--564.

\bibitem[Aitchison, 1994]{Aitchison_1994}
Aitchison, J. (1994).
\newblock Principles of compositional data analysis.
\newblock {\em Multivariate Analysis and Its Applications}, 24:73--81.

\bibitem[Aitchison, 2003]{aitchison2003the}
Aitchison, J. (2003).
\newblock {\em The statistical analysis of compositional data}.
\newblock Blackburn Press, Caldwell, NJ.

\bibitem[Aitchison and Egozcue, 2005]{Aitchison_2005}
Aitchison, J. and Egozcue, J.~J. (2005).
\newblock Compositional data analysis: Where are we and where should we be
  heading?
\newblock {\em Mathematical Geology}, 37(7):829--850.

\bibitem[Avalos et~al., 2018]{Avalos_2018}
Avalos, M., Nock, R., Ong, C.~S., Rouar, J., and Sun, K. (2018).
\newblock Representation learning of compositional data.
\newblock In Bengio, S., Wallach, H., Larochelle, H., Grauman, K.,
  Cesa-Bianchi, N., and Garnett, R., editors, {\em Advances in Neural
  Information Processing Systems}, volume~31, pages 6679--6689. Curran
  Associates, Inc.

\bibitem[Bergeron-Boucher et~al., 2017]{Bergeron_Boucher_2017}
Bergeron-Boucher, M.-P., Canudas-Romo, V., Oeppen, J., and Vaupel, J.~W.
  (2017).
\newblock Coherent forecasts of mortality with compositional data analysis.
\newblock {\em Demographic Research}, 37:527--566.

\bibitem[Chayes, 1960]{Chayes_1960}
Chayes, F. (1960).
\newblock On correlation between variables of constant sum.
\newblock {\em Journal of Geophysical Research}, 65(12):4185--4193.

\bibitem[Collins et~al., 2002]{Collins_2002}
Collins, M., Dasgupta, S., and Schapire, R.~E. (2002).
\newblock A generalization of principal components analysis to the exponential
  family.
\newblock In Dietterich, T., Becker, S., and Ghahramani, Z., editors, {\em
  Advances in Neural Information Processing Systems}, volume~14. MIT Press.

\bibitem[Denuit et~al., 2019]{Denuit_2019}
Denuit, M., Guillen, M., and Trufin, J. (2019).
\newblock Multivariate credibility modelling for usage-based motor insurance
  pricing with behavioural data.
\newblock {\em Annals of Actuarial Science}, 13(2):378--399.

\bibitem[Frees, 2009]{frees2009}
Frees, E.~W. (2009).
\newblock {\em Regression Modeling with Actuarial and Financial Applications}.
\newblock Cambridge University Press, Cambridge, UK.

\bibitem[Gan and Valdez, 2018]{gan2018}
Gan, G. and Valdez, E. (2018).
\newblock {\em Actuarial Statistics with R: Theory and Case Studies}.
\newblock ACTEX Learning.

\bibitem[Guillen et~al., 2019]{Guillen_2019}
Guillen, M., Nielsen, J.~P., Ayuso, M., and Pérez-Marín, A.~M. (2019).
\newblock The use of telematics devices to improve automobile insurance rates.
\newblock {\em Risk Analysis}, 39(3):662--672.

\bibitem[Guillen et~al., 2020]{Guillen_2020}
Guillen, M., Nielsen, J.~P., Pérez-Marín, A.~M., and Elpidorou, V. (2020).
\newblock Can automobile insurance telematics predict the risk of near-miss
  events?
\newblock {\em North American Actuarial Journal}, 24(1):141--152.

\bibitem[Pawlowsky-Glahn et~al., 2015]{pawlowsky2015modelling}
Pawlowsky-Glahn, V., Egozcue, J.~J., and Tolosana-Delgado, R. (2015).
\newblock {\em Modelling and analysis of compositional data}.
\newblock John Wiley \& Sons, Hoboken, NJ.

\bibitem[Pearson, 1897]{Pearson_1897}
Pearson, K. (1897).
\newblock Mathematical contributions to the theory of evolution. on a form of
  spurious correlation which may arise when indices are used in the measurement
  of organs.
\newblock {\em Proceedings of the Royal Society of London},
  60(359-367):489--498.

\bibitem[Pesantez-Narvaez et~al., 2019]{Pesantez_Narvaez_2019}
Pesantez-Narvaez, J., Guillen, M., and Alcañiz, M. (2019).
\newblock Predicting motor insurance claims using telematics data—xgboost
  versus logistic regression.
\newblock {\em Risks}, 7(2):70.

\bibitem[So et~al., 2021]{So_2021}
So, B., Boucher, J.-P., and Valdez, E.~A. (2021).
\newblock Cost-sensitive multi-class {A}da{B}oost for understanding driving
  behavior based on telematics.
\newblock {\em ASTIN Bulletin: The Journal of the International Actuarial
  Association}, 51(3):719--751.

\bibitem[Tipping and Bishop, 1999]{Tipping_1999}
Tipping, M.~E. and Bishop, C.~M. (1999).
\newblock Probabilistic principal component analysis.
\newblock {\em Journal of the Royal Statistical Society: Series B (Statistical
  Methodology),}, 61(3):611--622.

\bibitem[Verbelen et~al., 2018]{Verbelen_2018}
Verbelen, R., Antonio, K., and Claeskens, G. (2018).
\newblock Unravelling the predictive power of telematics data in car insurance
  pricing.
\newblock {\em Journal of the Royal Statistical Society: Series C (Applied
  Statistics)}, 67(5):1275--1304.

\end{thebibliography}

\appendix

\end{document}